\documentclass[a4paper,12pt]{article}

\pdfoutput=1 

\usepackage[hmargin=.7in,vmargin=1.1in]{geometry}
\usepackage{indentfirst}
\usepackage[mathcal]{euscript}
\usepackage{amsfonts}
\usepackage{mathrsfs}
\usepackage{amsmath}
\usepackage{amssymb}
\usepackage{authblk}
\usepackage{cite}
\usepackage{xcolor}
\usepackage{mathtools}
\usepackage{tensor}
\usepackage{physics}
\usepackage{graphicx}
\usepackage{bm}
\usepackage{upgreek}
\usepackage{braket}
\usepackage{color,soul}
\usepackage{csquotes}
\usepackage{caption}
\usepackage{subcaption}
\usepackage{orcidlink}

\hypersetup{
            colorlinks,
            linkcolor=[rgb]{0,0.3,0.6}, 
            citecolor=[rgb]{0,0.3,0.6}, 
            urlcolor=[rgb]{0,0.3,0.6}
           }

\def\be{\begin{eqnarray}}
\def\ee{\end{eqnarray}}

\begin{document}

\title{\Large\textbf{Stellar modeling within regularized 4D Einstein-Gauss-Bonnet gravity in light of current astrophysical constraints}}

\author[a]{
Grigorios Panotopoulos  \orcidlink{0000-0002-7647-4072}
\thanks{e-mail: 
\href{mailto:grigorios.panotopoulos@ufrontera.cl}
{\nolinkurl{grigorios.panotopoulos@ufrontera.cl}}}
}

\author[b c]{
\'Angel Rinc\'on {\orcidlink{0000-0001-8069-9162}} \thanks{e-mail: 
\href{mailto:angel.rincon@physics.slu.cz}
{\nolinkurl{angel.rincon@physics.slu.cz}}}
}

\author[d]{
Ilidio Lopes  \orcidlink{0000-0002-5011-9195}
\thanks{e-mail: 
\href{mailto:ilidio.lopes@tecnico.ulisboa.pt}
{\nolinkurl{ilidio.lopes@tecnico.ulisboa.pt}}}
}

\affil[a]{\normalsize{\em Departamento de Ciencias F{\'i}sicas, Universidad de la Frontera, Casilla 54-D, 4811186 Temuco, Chile.}}

\affil[b]{\normalsize{\em Research Centre for Theoretical Physics and Astrophysics, Institute of Physics, Silesian University in Opava,
Bezručovo náměstí 13, CZ-74601 Opava, Czech Republic.}}

\affil[c]{\normalsize{\em
Instituto Universitario de Matemática Pura y Aplicada,
Universitat Polit\`ecnica de Val\`encia, Valencia 46022, Spain.}
}

\affil[d]{\normalsize{\em Centro de Astrof{\'i}sica e Gravita{\c c}{\~a}o, Departamento de F{\'i}sica, Instituto Superior T{\'e}cnico-IST, Universidade de Lisboa-UL, Av. Rovisco Pais, 1049-001 Lisboa, Portugal.}}

\date{ }

\maketitle

\begin{abstract}
\smallskip\noindent	
In this study we obtain interior solutions and investigate structural properties of isotropic compact stars in the framework of four-dimensional regularized Einstein-Gauss-Bonnet (4DEGB) gravity. For stellar matter content, we adopt a widely used quark-matter model that approximates a realistic equation of state (EoS). By numerically integrating the modified Tolman-Oppenheimer-Volkoff equations, we obtain interior solutions for static, spherically symmetric fluid spheres. The resulting sequences are compared directly with the predictions of General Relativity (GR). Our analysis focuses on three diagnostic indicators:  (i) the mass-radius profiles under GR and three representative choices of the Gauss-Bonnet coupling; (ii) the stellar compactness factor, $C \equiv M/R$; and (iii) the relation between stellar mass and central energy density. Recent observational studies suggest that the maximum masses inferred from the mass-radius relation may be larger than previously expected. To address this, we include a comparative set of constraints from multi-messenger astrophysical observations, including gravitational-wave event GW190814, as well as X-ray measurements from NICER for PSR~J0740+6620 and PSR~J0030+0451. These data provide stringent, astrophysically grounded tests of the viability of the models discussed here. Our results indicate that compact stars within 4DEGB gravity are systematically less compact and achieve moderately higher maximum masses compared to the GR case. This trend is consistent with recent theoretical analyses of compact stars in higher-curvature gravity theories and with constraints from multi-messenger astrophysics. Together, these findings suggest that regularized Gauss-Bonnet corrections constitute a plausible extension of GR in the strong-field regime.  
\end{abstract}

{\bf{Keywords:}} 
Theories of gravity other than GR,
Relativistic stars,
Stellar composition,
Equation-of-state.

\tableofcontents

\newpage

\section{Introduction}
\label{sec:intr}

\smallskip\noindent

Despite the outstanding success of Einstein's General Relativity (GR) \cite{Einstein:1916vd}, alternative theories of gravity continue to attract sustained interest in theoretical physics \cite{2012PhR...513....1C}. These efforts are motivated by the need to resolve cosmological tensions \cite{Bouche:2022qcv,McCarthy:2023ism,Pogosian:2021mcs,Alvarez:2020xmk}, advance a consistent quantum theory of gravity, alleviate spacetime singularities, and offer frameworks that permit rigorous tests of GR in the strong-field regime.
Such investigations are particularly relevant for compact astrophysical objects, where recent observational advances now allow for strong-field phenomena to be probed with unprecedented precision \cite{2019PhRvL.123a1102A}.
 
\smallskip\noindent

Among the proposed extensions of GR, higher-order curvature theories (HCTs) represent a widely studied class \cite{Chakraborty:2016ydo,Capozziello:2011et,Capozziello:2007ec}. They generalize the Einstein field equations beyond the linear dependence of curvature on stress-energy by introducing polynomial contributions from higher-order curvature terms. These modifications enhance theoretical flexibility and may produce observable deviations that can be examined with astrophysical data. A prominent case is four-dimensional Einstein-Gauss-Bonnet gravity (4DEGB), obtained by dimensional regularization of the Gauss-Bonnet action \cite{2020PhRvL.124h1301G,2020PhLB..80535468F,2020JHEP...07..027H}. Unlike the trivial Gauss-Bonnet term in four dimensions, the regularized approach preserves quadratic curvature effects, enabling their study in relativistic stellar astrophysics.

\smallskip\noindent

Independent theoretical motivations arise from quantum gravity \cite{ahmed2017} and cosmological frameworks involving dark energy, dark matter, and inflation \cite{Bueno:2016xff}, which all suggest that HCTs may play a vital role in fundamental physics. Lovelock theories \cite{lovelock1971} are of particular interest, since they retain second-order field equations despite non-linear curvature couplings. However, their contributions vanish for $D \leq 4$, limiting their applicability in four-dimensional space-times. This issue has been overcome by the construction of a consistent regularized formulation of 4DEGB gravity \cite{2020PhRvL.124h1301G,2020PhRvD.102b4025F}, where quadratic terms maintain dynamical significance and allow systematic applications to compact objects.

\smallskip\noindent

Relativistic stars provide an effective natural laboratory to test strong-field gravity. Key observables such as the mass-radius relation, compactness, and tidal deformability can be directly compared with constraints from radio pulsar timing, X-ray measurements, and gravitational-wave observations \cite{LIGOScientific:2020zkf,Miller:2021qha,Riley:2021pdl}. The importance of these tests has been reinforced by multimessenger detections of binary neutron star mergers \cite{2018PhRvL.121i1102D}. Recent studies \cite{2021JCAP...05..024D,2025PhRvD.111f4071S} demonstrate that Gauss-Bonnet corrections induce measurable modifications to the stellar mass-radius sequences and stability properties, while remaining compatible with present astrophysical data.


\smallskip\noindent

Last but not least, it is essential to point out that the study of compact relativistic stars within the framework of regularized Gauss-Bonnet gravity is motivated by both i) observational tensions and ii) theoretical developments in gravitational physics. 
First, from an observational perspective, recent high-precision measurements from NICER, LIGO/Virgo, and other multi-messenger astrophysical probes have begun to explore the strong-field regime of gravity, where deviations from general relativity may become significant.
In particular, the detection of massive neutron stars exceeding two solar masses (such as PSR J0740+6620 \cite{Miller:2021qha,Riley:2021pdl}, for instance), challenges many soft equations of state and suggests the need for additional gravitational support, precisely the arena in which RGB gravity can naturally provide through higher-curvature corrections. 
Furthermore, constraints on tidal deformability from gravitational wave events (like GW170817 \cite{LIGOScientific:2017vwq}), as well as surface redshift measurements from X-ray spectroscopy, offer the change to test the modified internal structure and compactness predictions of RGB gravity. 
Second, from an theoretical point of view, RGB gravity appears as a well-motivated and self-consistent extension of GR, inspired by the low-energy limit of string theory and higher-dimensional models, and is constructed to evade the Lovelock theorem’s triviality in four dimensions via regularization techniques. 
Unlike many higher-order theories, RGB gravity maintains a well-posed initial value formulation and avoids ghost instabilities, making it suitable for astrophysical modeling. It also allows for non-perturbative phenomena such as spontaneous scalarization and the emergence of new equilibrium branches, enriching the landscape of possible stellar configurations. 
As such, compact stars serve as natural laboratories for probing the viability of RGB gravity, offering a unique opportunity to confront fundamental theories of gravity with empirical data.

\smallskip\noindent

This work aims to determine whether regularized Gauss-Bonnet corrections, characterised by the coupling parameter $\alpha$, result in stellar models that not only extend the classical predictions of GR but also remain consistent with observational constraints. Our analysis therefore provides further phenomenological insight into the viability of higher-curvature theories in describing strong-field physics within the astrophysical domain.

\smallskip\noindent

The structure of our paper is as follows. In Section \ref{sec:back}, we review the theoretical setup of 4DEGB gravity, and discuss the role of the Gauss-Bonnet term in four dimensions. Next, we present the generalized Tolman-Oppenheimer-Volkoff (TOV) equations for a realistic perfect-fluid source in Section \ref{sec:TOV}. We also discuss the boundary and initial conditions necessary for stellar equilibrium. In Section \ref{sec:pert}, we introduce the equations of state under consideration, provide the motivation for their selection, and highlight their relevance to realistic compact star modeling. Section \ref{sec:meth} contains the numerical results, including stellar mass-radius profiles, compactness factors, and the dependence of stellar mass on central density. Finally, in Section \ref{sec:disc}, we analyze the implications of these results and summarize our conclusions, with emphasis on future directions such as tidal deformability studies and rotating stellar configurations.

\section{Towards a Consistent Regularization of Gauss-Bonnet Gravity}
\label{sec:back}

\subsection{First attempts}

\smallskip\noindent

Let us start by considering the Gauss-Bonnet (GB) contribution to the gravitational action, i.e.,
\begin{equation}\label{eq:dgbaction}
S_{D}^{G B} = \alpha \int d^{D} x \sqrt{-g} \left[R^{\mu\nu\rho\tau}R_{\mu\nu\rho\tau} - 4 R^{\mu\nu}R_{\mu\nu} + R^2 \right]
\equiv \alpha \int d^{D} x \sqrt{-g} \mathcal{G},
\end{equation}
(where  $R_{\mu\nu\rho\tau}$ is the usual Riemann curvature tensor in $D$ dimensions and $\mathcal{G}$ is the GB term)
and which becomes the integral of a total derivative in $D=4$, meaning that this term does not 
contribute to the gravitational dynamics in less than five dimensions. It is usually referred to as a ``topological term" because it does not affect the equations of motion.  
Nevertheless, a non-trivial redefinition of the Gauss-Bonnet coupling constant $\alpha$ was identified several years ago (see \cite{2020PhRvL.124h1301G} for details), which could be an alternative way of accounting for the impact of the Gauss-Bonnet term in four dimensions. Thus, the redefinition
\begin{equation}
\label{eq:alpharescale}
\lim_{D \to 4} (D-4)\;\; \alpha \rightarrow \alpha.
\end{equation}
The idea was implemented in many occasions, to investigate the impact of the GB corrections on four-dimensional backgrounds (see \cite{2020PhRvL.124h1301G,kobayashi2020,kumar2022,2020PhLB..80535468F,kumar2022_2,malafarina2020,charmousis2022,ghosh2020} and references therein). Roughly speaking, the corrections encoded imprints of the quadratic curvature effects of their $D > 4$ counterparts.   
 
\smallskip\noindent

Subsequent critiques of this approach \cite{gurses2020,ai2020,shu2020} highlighted that the existence of a limiting solution does not necessarily guarantee a well-defined four-dimensional (4D) theory with field equations that admit such a solution. This issue was promptly resolved by demonstrating that the $D \to 4$ limit, as outlined in \eqref{eq:alpharescale}, can be consistently applied to the gravitational action \cite{2020JHEP...07..027H,2020PhRvD.102h4005C}.
This method generalizes an earlier technique used to derive the $D \to 2$ limit of General Relativity \cite{Mann:1992ar}.
 
\subsection{Regularized Gauss-Bonnet action}

\smallskip\noindent

Taking into account the re-scaling of the Gauss-Bonnet coupling constant, combined to the inclusion of a scalar field $\phi$ coupled to the GB term, the resulting theory is none other than a scalar-tensor modification of GR. This resulting theory belongs to the Horndeski class, representing the most general scalar-tensor theories yielding second-order field equations, thereby avoiding Ostrogradsky instabilities \cite{2020JHEP...07..027H}.	
The action for the four-dimensional Einstein-Gauss-Bonnet gravity (4DEGB) is given by \cite{Gammon:2023uss}:
\begin{equation}
\begin{aligned}
\mathcal{S}[g_{\mu \nu},\phi] &\equiv   \frac{1}{2\kappa} \left(S^{GR} + S_{4}^{GB}\right) + \mathcal{S}_m
\\
\mathcal{S}[g_{\mu \nu},\phi]&= \frac{1}{2\kappa}\int \mathrm{d}^{4} x \sqrt{-g}\left[R + 
\alpha\left\{  \phi \mathcal{G}+4 G_{\mu \nu} \nabla^\mu \phi \nabla^\nu \phi-4(\nabla \phi)^2 \square \phi+2(\nabla \phi)^4 \right\}\right] + \mathcal{S}_m
\end{aligned}
\label{4DEGB}
\end{equation}
whose action is given by GB term plus the usual Einstein-Hilbert term. In addition, $\kappa \equiv 8 \pi G$ is the gravitational constant, $G$ is the 4D Newton's coupling and, finally, $\mathcal{S}_m$ is the matter contribution.
It is noteworthy that this novel gravitational action has been demonstrated to serve as a compelling phenomenological alternative to General Relativity (GR) \cite{Clifton:2020xhc}. 

\smallskip\noindent

The modified Gauss-Bonnet (GB) framework provides a unique opportunity to study the impact of higher-curvature corrections on compact astrophysical objects, including quark stars (QSs)  \cite{2021JCAP...05..024D,banerjee_2021_quark,Gammon:2023uss,Pretel:2025roz}, electrically charged quark stars \cite{Pretel2022EPJC,Gammon2025arxiv}, and neutron stars (NSs) \cite{2021JCAP...05..024D,2025PhRvD.111f4071S}. 
Our findings confirm that incorporating the GB term, characterized by its coupling constant $\alpha$, introduces novel phenomenological signatures that can  modify  the mass-radius relationships and stability properties of relativistic stars.

\subsection{Field Equations: 4D Einstein-Gauss-Bonnet Gravity}

\smallskip\noindent

In order to obtain the field equations of the theory, let us start by applying the variational principle to the action \eqref{4DEGB}. Thus, variation with respect to the scalar $\phi$ yields the following equation:
\begin{equation}\label{eq:eomscalar}
\begin{aligned}
\mathcal{E}_{\phi}=&-\mathcal{G}+8 G^{\mu \nu} \nabla_{\nu} \nabla_{\mu} \phi+8 R^{\mu \nu} \nabla_{\mu} \phi \nabla_{\nu} \phi-8(\square \phi)^{2}+8(\nabla \phi)^{2} \square \phi+16 \nabla^{a} \phi \nabla^{\nu} \phi \nabla_{\nu} \nabla_{\mu} \phi \\
&\qquad +8 \nabla_{\nu} \nabla_{\mu} \phi \nabla^{\nu} \nabla^{\mu} \phi \\
=& \; 0.
\end{aligned}
\end{equation}
Next, variational principle with respect to the metric tensor yields the gravitational field equations
\begin{equation}\label{eq:eommetric}
\begin{aligned}
\mathcal{E}_{\mu \nu} &=\Lambda g_{\mu \nu}+G_{\mu \nu}+\alpha\left[\phi H_{\mu \nu}-2 R\left[\left(\nabla_{\mu} \phi\right)\left(\nabla_{\nu} \phi\right)+\nabla_{\nu} \nabla_{\mu} \phi\right]+8 R_{(\mu}^{\sigma} \nabla_{\nu)} \nabla_{\sigma} \phi+8 R_{(\mu}^{\sigma}\left(\nabla_{\nu)} \phi\right)\left(\nabla_{\sigma} \phi\right)\right.\\
&-2 G_{\mu \nu}\left[(\nabla \phi)^{2}+2 \square \phi\right]-4\left[\left(\nabla_{\mu} \phi\right)\left(\nabla_{\nu} \phi\right)+\nabla_{\nu} \nabla_{\mu} \phi\right] \square \phi-\left[g_{\mu \nu}(\nabla \phi)^{2}-4\left(\nabla_{\mu} \phi\right)\left(\nabla_{\nu} \phi\right)\right](\nabla \phi)^{2} \\
&+8\left(\nabla_{(\mu} \phi\right)\left(\nabla_{\nu)} \nabla_{\sigma} \phi\right) \nabla^{\sigma} \phi-4 g_{\mu \nu} R^{\sigma \rho}\left[\nabla_{\sigma} \nabla_{\rho} \phi+\left(\nabla_{\sigma} \phi\right)\left(\nabla_{\rho} \phi\right)\right]+2 g_{\mu \nu}(\square \phi)^{2} \\
& -4 g_{\mu \nu}\left(\nabla^{\sigma} \phi\right)\left(\nabla^{\rho} \phi\right)\left(\nabla_{\sigma} \nabla_{\rho} \phi\right)+4\left(\nabla_{\sigma} \nabla_{\nu} \phi\right)\left(\nabla^{\sigma} \nabla_{\mu} \phi\right) \\ 
&\left. -2 g_{\mu \nu}\left(\nabla_{\sigma} \nabla_{\rho} \phi\right)\left(\nabla^{\sigma} \nabla^{\rho} \phi\right)
+4 R_{\mu \nu \sigma \rho}\left[\left(\nabla^{\sigma} \phi\right)\left(\nabla^{\rho} \phi\right)+\nabla^{\rho} \nabla^{\sigma} \phi\right] \right]\\
\mathcal{E}_{\mu \nu} &=\;T_{\mu \nu}
\end{aligned}
\end{equation}
where 
\begin{equation}\label{eq:gbtensor}
    \begin{aligned}
    H_{\mu \nu}=2\Big[R R_{\mu \nu}-2 R_{\mu \alpha \nu \beta} R^{\alpha \beta}+R_{\mu \alpha \beta \sigma} R_{\nu}^{\alpha \beta \sigma}-2 R_{\mu \alpha} R_{\nu}^{\alpha} -\frac{1}{4} g_{\mu \nu}\mathcal{G}
\Big]
    \end{aligned}
\end{equation}
is the so-called Gauss-Bonnet tensor $ H_{\mu \nu}$. 
The energy-momentum tensor satisfies the following equation:
\begin{equation}\label{eq:fieldeqntrace}
\kappa g^{\mu \nu}T_{\mu \nu}=g^{\mu \nu} \mathcal{E}_{\mu \nu}+\frac{\alpha}{2} \mathcal{E}_{\phi}=4 \Lambda-R-\frac{\alpha}{2} \mathcal{G}
\end{equation}
which helps in confirming whether or not prior solutions generated are even possible solutions to the theory. The prior solutions in question are those obtained using the Glavan/Lin \cite{2020PhRvL.124h1301G}  rescaling approach from the $D\rightarrow 4$ limit, whose consistency was subsequently established by the regularization procedures of Hennigar et al.\cite{2020JHEP...07..027H} and Fernandes et al. \cite{2020PhRvD.102b4025F}.

\section{Fluid spheres made of isotropic matter: Modified structure equations}
\label{sec:TOV}

\smallskip\noindent

Although the Tolman-Oppenheimer-Volkoff (TOV) equations for stellar structure are well known in GR \cite{Tolman:1939jz,Oppenheimer:1939ne}, they should be modified to account for corrections from regularized Gauss-Bonnet gravity (i.e. higher-derivative terms). Let us consider static, spherically symmetric compact stars, for which the line element is given by
\begin{eqnarray}\label{metric}
  \mathrm{d}s^2 = - e^{2\Phi(r)}\mathrm{d}t^2 + e^{2\Psi(r)}\mathrm{d}r^2 + r^2\mathrm{d}\Omega^2,
\end{eqnarray}
with $\{\Phi(r),\Psi(r)\}$ being unknown metric functions to be obtained after solving the system of equations. The isotropic matter content is represented by 
\begin{eqnarray}\label{EMT}
T_{\mu \nu} \equiv (\rho + p)u_\mu u_\nu + pg_{\mu \nu}, 
\end{eqnarray}
where $\rho$ and $p$ are the energy density and the radial pressure, respectively.
To obtain the effective TOV equation, let us take advantage of the 00 and 11 components of the field equations:
\begin{align}
    \frac{2}{r}\left[ 1+ \frac{2\alpha(1-e^{-2\Psi})}{r^2} \right]\frac{d\Psi}{dr} &= e^{2\Psi} \left[ \kappa  \rho - \frac{1- e^{-2\Psi}}{r^2}\left( 1- \frac{\alpha(1- e^{-2\Psi})}{r^2} \right) \right] , \label{FEq1}  \\
    \frac{2}{r}\left[ 1+ \frac{2\alpha(1-e^{-2\Psi})}{r^2} \right]\frac{d\Phi}{dr} &= e^{2\Psi} \left[ \kappa  p + \frac{1- e^{-2\Psi}}{r^2}\left( 1- \frac{\alpha(1- e^{-2\Psi})}{r^2} \right) \right] ,\label{FEq2}
\end{align}
where in natural geometric units $(G=1=c)$, the gravitational constant takes the value $\kappa = 8 \pi$.
The covariant conservation of the energy-momentum tensor for the isotropic case provides the following fluid equation:
\begin{equation}\label{ConsevEq}
    \frac{\mathrm{d}p_r}{\mathrm{d}r} = -(\rho+p) \frac{\mathrm{d}\Phi}{\mathrm{d}r}.
\end{equation}
The relation between the mass function $m(r)$ and metric potential $\Psi(r)$ is given by \cite{2020JHEP...07..027H, Gammon:2023uss} 
\begin{equation}\label{PsiEq}
    e^{-2\Psi} = 1+ \frac{r^2}{2\alpha}\bigg( 1-  \mathcal{A}(r;\alpha) \bigg) ,
\end{equation}
where for convenience, we have defined the function $\mathcal{A}(r;\alpha)$ as follows:
\begin{align}
    \mathcal{A}(r;\alpha) &\equiv \sqrt{1+ \frac{8\alpha m(r)}{r^3}},
\end{align}
and where as usual $m(r)$ is the enclosed gravitational mass within the radial coordinate $r$. As expected, when the Gauss-Bonnet coupling constant is small, the Schwarzschild geometry \cite{Schwarzschild:1916uq} is recovered, i.e., 
\begin{align}
    e^{2\Psi(r)} & \approx 
    \Bigg(1- \frac{2m}{r} + \frac{4m^2}{r^{4}}\alpha + \mathcal{O}(\alpha^2)
    \Bigg)^{-1}.
\end{align}
Finally, utilizing Eqs.~\eqref{FEq1} and \eqref{ConsevEq} we obtain respectively
\begin{align}
    \frac{\mathrm{d}m}{\mathrm{d}r} &= 4\pi r^2\rho ,  \label{TOV1}  
    \\
    \frac{\mathrm{d}p}{\mathrm{d}r} &= (\rho+ p) \frac{\left[ 2\alpha m + r^3 (1- \mathcal{A}- 8\pi\alpha p) \right]}{r^2\mathcal{A}\left( r^2+ 2\alpha - r^2\mathcal{A} \right)} .  
    \label{TOV2} 
\end{align}

\smallskip\noindent
The last two differential equations \eqref{TOV1} and \eqref{TOV2} are the modified TOV equations within 4DEGB gravity  \cite{Gammon:2023uss}. 
With these equations at hand, the system can be closed by supplementing the set of differential equations with a suitable equation of state (EoS), which will be discussed in the next Section. At this stage, it is important to note that our focus lies on physically realistic scenarios; therefore, in the following we shall restrict ourselves to well-motivated EoSs derived from the particle physics we know.

\smallskip\noindent

Numerical solutions to the problem require specific boundary conditions. Thus, we are interested in the region defined by:
i) the stellar centre ($r = 0$) and ii) the stellar surface ($r = R$). Firstly, at the centre, the initial conditions are as follows: 
\begin{equation}
m(0)=0, \; \; \; \; \; \; p(0)=p_c,
\end{equation}
namely the enclosed mass is zero at the origin, while the pressure takes an arbitrary value that may vary.
Next, at the surface of the star we impose the matching conditions:
\begin{equation}
p(R)=0, \; \; \; \; \; \; m(R)=M, \; \; \; \; \; \; e^{2 \Phi(R)}=F(R),
\end{equation}
with $F(r)$ being the metric function of the exterior vacuum geometry
\begin{equation}
\label{eq:exterior}
F(r) = 1+ \frac{r^2}{2\alpha} \left( 1- \sqrt{1 + \frac{8\alpha M}{r^3}} \right),
\end{equation}
which generalizes the usual Schwarzschild geometry of GR \cite{Schwarzschild:1916uq}.
Integration finishes when the pressure drops to zero, which defines the stellar radius $R$. The stellar mass of the compact star is then given by $M \equiv m(R)$.
Having clarified the boundary conditions, it is worth noting that generating several curves on the $M-R$ diagram (to explore the structural properties of stars) requires consistent variation of the central density, so that the standard mass-to-radius profile for this type of relativistic star can be recovered.

\section{Matter content: Equation of state}
\label{sec:pert}

\smallskip\noindent

In this section, in light of the theoretical framework and field equations for compact stars within the context of regularized Gauss-Bonnet gravity in four dimensions, we investigate one concrete equation of state to demonstrate how the mass-to-radius relation is affected by the inclusion of higher-derivative gravity corrections. Furthermore, we discuss the differences between the stellar solutions arising from the two models considered in this study.

\smallskip\noindent

The equation of state (EoS) plays a crucial role in compact star physics by describing matter at extreme densities and determining stellar mass and radius, and stability of the configuration. More precisely, the EoS is a non-trivial relationship connecting thermodynamic variables that specify the state of a physical system. Broadly speaking, the EoS can be mathematically expressed as an expansion of pressure in powers of density, where the coefficients in the series encode deviations from the simplest scenario and can be derived from the underlying elementary interactions.
Consequently, the EoS encapsulates essential dynamical information, enabling a suitable choice to link measurable macroscopic quantities with the forces acting between the constituents of the system at microscopic level.

\smallskip\noindent

Now, let us concretize the EoS to be used for the present case. Strange quark stars are based on the seminal works from the 70s and 80s \cite{Itoh:1970uw,Bodmer:1971we,Witten:1984rs,Terazawa:1989iw}, where it was proposed that quark matter is by assumption absolutely stable, and as such it may be the true ground state of Quantum Chromodynamics (QCD). According to this idea, up, down and strange quarks in weak equilibrium, in the stellar interior become effectively massless as compared to the associated chemical potential at very large densities, forming Cooper pairs with a common Fermi momentum. Since those pairs are electrically neutral, electrons cannot be present in this superfluid ground state \cite{Rajagopal:2000ff}, dubbed color-flavor-locked (CFL) phase.

\smallskip\noindent

For CFL strange stars, the equation of state of quark matter may be obtained within the framework of the MIT bag model \cite{Chodos:1974pn,Chodos:1974je,Johnson1975}, although now due to QCD superconductivity effects the linear EoS of the simplest version receives corrections of order $(\Delta/\mu)^2$, which is around a few percent for typical values of the color superconducting energy gap ($\Delta \sim 0-150$ MeV) and the baryon chemical potential ($\mu \sim 300-400$ MeV). From all the possible viable models, we shall adopt here the model CFL 9 \cite{VasquezFlores:2017uor},  for which the numerical values of the 3 parameters are as follows:
\begin{equation}
B = 80 \ \frac{\text{MeV}}{\text{fm}^3}, \; \; \; \; \Delta = 150 \ \text{MeV},  \; \; \; \; m_s = 0 \ \text{MeV},
\end{equation}
with $m_s$ being the mass of the s quark, and $B$ being the bag constant.
To order $\Delta^{2}$ and  $m^{2}_{s}$ the pressure and energy density can be written as \cite{lugones2002}:
\begin{equation}\label{PB}
p=\frac{3\mu^{4}}{4\pi^{2}}+\frac{9 a \mu^{2}}{2\pi^{2}} - B,
\end{equation}
\begin{equation}\label{energiaB}
\rho = \frac{9\mu^{4}}{4\pi^{2}}+\frac{9 a \mu^{2}}{2\pi^{2}} + B,
\end{equation}
where
\begin{equation}\label{alfa}
a= -\frac{m^{2}_{s}}{6}+\frac{2\Delta^{2}}{3}.
\end{equation}
From the above expressions, we can obtain an analytic expression for $\rho = \rho(p)$: 
\begin{equation}\label{energiaC}
\rho = 3p+4B-\frac{9 a \mu^{2}}{\pi^{2}},
\end{equation}
with
\begin{equation}\label{mu2}
\mu^{2}= -3 a +\bigg[\frac{4}{3}\pi^{2}(B+p)+9 a^{2}\bigg]^{1/2},
\end{equation}

\smallskip\noindent 
to finally obtain
\begin{equation}
\rho = 3p+4B-\frac{9 a}{\pi^{2}} 
\Bigg[-3 a +\sqrt{\frac{4}{3}\pi^{2}(B+p)+9 a^{2}}
\Bigg].
\end{equation}


\begin{figure*}[h]
\centering
\includegraphics[width=0.90\textwidth]{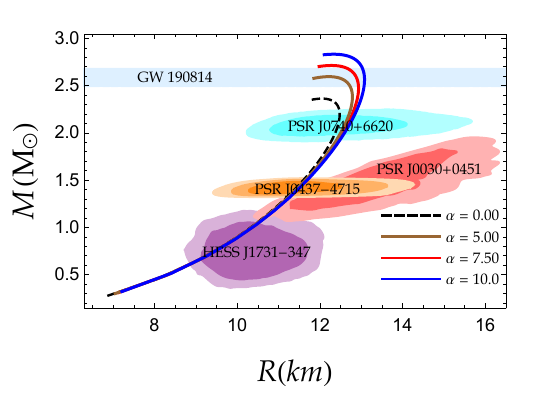}
\caption{
\textcolor{black}{
Stellar mass in solar masses versus stellar radius in km assuming $\alpha=5.00$ (solid brown line), $\alpha=7.50$ (solid red line), and $\alpha=10.0$ (solid blue line). 
For comparison reasons, we have also shown the mass-to-radius relationship for the case of GR $(\alpha=0)$ (dashed black line). 
We have included one horizontal strip around 2.5 solar masses. The light-blue band corresponds to the GW event 190814 \cite{LIGOScientific:2020zkf}.
In addition, there are four regions corresponding to:
a) the light HESS compact object (purple region) \cite{Doroshenko:2022nwp},
b) the pulsar J0740+6620 (cyan region) \cite{Miller:2021qha,Riley:2021pdl}, 
c) the pulsar J0030+0451 (red region), \cite{Miller:2019cac}, 
and 
d) the light HESS compact object (purple region) \cite{Choudhury:2024xbk}.
The intensity of the color represents 65$\%$, 90$\%$ and 99$\%$ 
confidence levels (CLs) from darker to lighter color, delineating the observationally allowed mass-radius parameter space for each compact object.}}
\label{fig:MR-profile1}
\end{figure*}


\begin{figure*}[h]
	\centering
	\includegraphics[width=0.495\textwidth]{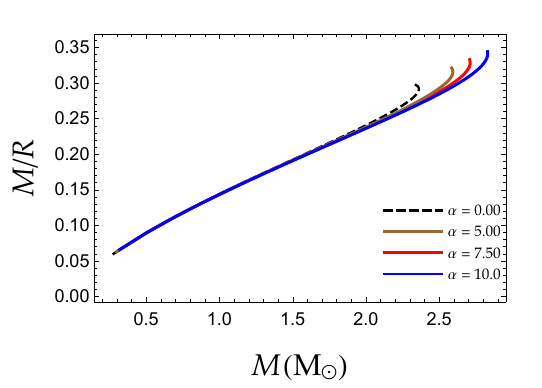}
	\includegraphics[width=0.495\textwidth]{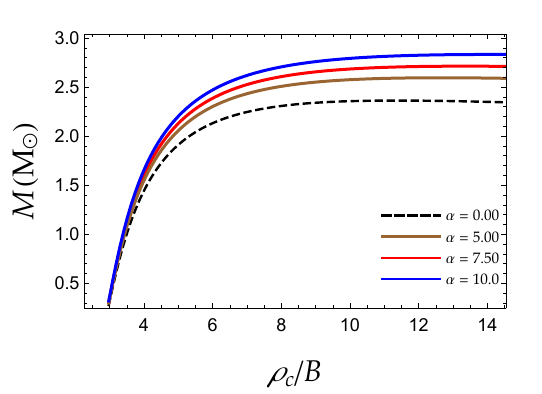}
	\caption{
		\textcolor{black}{
			{\bf{Left panel:}} Factor of compactness, $C\equiv M/R$, versus stellar mass (in solar masses), considering the same meaning of the color code as in the previous figure.
			{\bf{Right panel:}} Stellar mass as a function of the (normalized) central energy density of quark matter, considering the same meaning of the color code as in the previous figure.
		}
	}
	\label{fig:MR-profile2}
\end{figure*}


\begin{table}[]
\centering
\caption{
Stellar mass, radius and factor of compactness at the highest stellar mass.
}
\label{tab:frequencies}
\begin{tabular}{@{}c|cccc@{}}
\hline
\hline
 & $\alpha = 0.0~(km)^2$  & $\alpha = 5.0~(km)^2$ & $\alpha=7.5~(km)^2$  & $\alpha=10.0~(km)^2$
\\
\hline
$M_{max}~(M_{\odot})$ & 2.36 & 2.59 & 2.71 & 2.83  \\
$R~(km)$  & 12.05 & 12.21 & 12.29 & 12.34  \\
$C=M_{max}/R$  & 0.29 & 0.32 & 0.33 & 0.34
\end{tabular}
\end{table}

\section{Numerical Solutions}
\label{sec:meth}

\smallskip\noindent

We present comprehensive numerical results from integrating the modified stellar structure equations within regularized four-dimensional Einstein-Gauss-Bonnet (4DEGB) gravity. The calculations employ a color-flavor locked quark matter equation-of-state, specifically the CFL9 model derived from the MIT bag framework with corrections of order $(\Delta/\mu)^2$ \cite{VasquezFlores:2017uor,lugones2002}. This semi-analytic treatment enables direct comparison between modified gravity effects and general relativity (GR) whilst maintaining control over microphysical parameters.

\smallskip\noindent

The numerical integration scheme solves the coupled differential equations \eqref{TOV1} and \eqref{TOV2} with appropriate boundary conditions: vanishing mass at the origin ($m(0)=0$) and specified central energy densities $\rho_c$ spanning the range $4B$ to $14B$, where $B$ represents the bag constant. Integration proceeds throughout the star from the center outwards, until the pressure vanishes, thereby defining the stellar surface radius $R$ and total gravitational mass $M=m(R)$. This procedure generates complete mass-radius sequences for each value of the Gauss-Bonnet coupling parameter $\alpha$.

\smallskip\noindent

Figure~\ref{fig:MR-profile1} illustrates the primary diagnostic: mass-radius relationships for four distinct cases comprising the GR limit ($\alpha=0$) and three positive coupling values ($\alpha=5.0, 7.5, 10.0$ in $(\text{km})^2$). 
It should be emphasized that recent studies discuss observational constraints on the coupling $\alpha$ (see, e.g., \cite{Gammon:2023uss,2020PhRvD.102h4005C,fernandes_2022_the}).
A physically acceptable range is $0 < \alpha < 10^{10} \ \text{m}^2$, with the lower bound derived from early universe cosmology and atomic nuclei, and the upper bound from LAGEOS satellite constraints.
When recent gravitational wave data are included, this range is further refined, yielding a more tightly constrained region of $0 < \alpha < 10 \ \text{(km)}^2$ (see \cite{fernandes_2022_the} for instance).
The theoretical curves demonstrate systematic trends with increasing $\alpha$, yielding configurations with marginally enhanced maximum masses and moderately larger radii at fixed mass. These deviations remain constrained, producing stellar models that are approximately 5--10 per cent less compact than their GR counterparts whilst maintaining astrophysical viability.

\smallskip\noindent

Current observational constraints provide crucial validation of our models. The sequences satisfy the fundamental two-solar-mass criterion established by PSR J0740+6620 \cite{Miller:2021qha,Riley:2021pdl} and PSR J1614--2230 \cite{Antoniadis:2013pzd}. Moreover, the computed curves encompass the parameter spaces determined by NICER measurements for PSR J0030+0451 \cite{Miller:2019cac} and PSR J0437--4715 \cite{Choudhury:2024xbk},  as well as the intriguing compact object in HESS J1731--347 \cite{Doroshenko:2022nwp}. The horizontal band near $M\simeq 2.5\,M_\odot$ corresponds to GW190814's secondary component, whose nature remains debated \cite{LIGOScientific:2020zkf}.

\smallskip\noindent

Figure~\ref{fig:MR-profile2} shows complementary diagnostics relevant to the analysis of stellar structure. The left panel illustrates the compactness parameter, $C \equiv M/R$, as a function of stellar mass. The sequences demonstrate that the inclusion of higher-curvature corrections reduces compactness at fixed mass. 
In 4DEGB gravity, the compactness bound differs from the classical GR Buchdahl limit of $C < 4/9$. As demonstrated by Chakraborty and Dadhich
\cite{2020PDU....3000658C}  and confirmed by  \cite{2025PhRvD.111d3034G,2024PhRvD.109b4026G},  positive values of the Gauss-Bonnet coupling $\alpha$ weaken the effective gravitational interaction in the strong-field regime, allowing stellar configurations to achieve higher compactness than permitted in GR. For certain equations of state and $\alpha$ values, stars can even exist with radii smaller than the usual Schwarzschild radius of GR, $R_S = 2GM/c^2$, though they remain outside their corresponding 4DEGB black hole horizons. 
The physical origin of this phenomenon lies in the modified horizon structure of 4DEGB gravity. For a black hole of mass $M$, the event horizon radius is obtained by solving $F(r_{H}) = 0$ in the exterior vacuum metric function given by Eq.~(\ref{eq:exterior}), i.e., setting
$ 1 + \frac{r_{H}^{2}}{2\alpha}\left(1 - \sqrt{1 + {8\alpha M}/{r_{H}^{3}}}\right) = 0,$
which yields the horizon radius $ r_{H} = M + \sqrt{M^{2} - \alpha} $ (in geometric units where $G = c = 1$). Since $\sqrt{M^{2} - \alpha} < M$ for $\alpha > 0$, this horizon radius is smaller than the usual Schwarzschild radius of GR, $R_{S} = 2M$, for all positive values of the coupling. Stellar configurations with radii in the interval $r_{H} < R < 2M$ are therefore permitted, representing objects more compact than the GR Schwarzschild limit while remaining outside the 4DEGB black hole horizon. This phenomenon has been demonstrated explicitly for quark stars by Gammon, Rourke, and Mann ~\cite{2024PhRvD.109b4026G,2025PhRvD.111d3034G}, and the theoretical foundation for modified compactness bounds in higher-curvature theories was established by Chakraborty and Dadhich \cite{2020PDU....3000658C}.
The exact modified bound depends on both $\alpha$ and the specific equation of state employed. The right panel presents the relation between stellar mass and central energy density. The curves exhibit the expected turning-point behavior, which marks the threshold of radial instability according to the Harrison-Zel’dovich criterion \cite{1965gtgc.book.....H,1971reas.book.....Z}. This stability property is consistent with previous analyses of compact stars in four-dimensional Einstein-Gauss-Bonnet gravity \cite{2025PhRvD.111f4071S}.

\smallskip\noindent

The region where $\mathrm{d}M/\mathrm{d}\rho_c>0$ (cf.\ Figure~\ref{fig:MR-profile2}, right panel) delineates the stable branch of stellar configurations, extending from low central densities up to the model's maximum mass. At central densities beyond this peak, a negative gradient in the mass-central density curve marks the transition to dynamical instability against radial perturbations. As the Gauss-Bonnet coupling parameter $\alpha$ increases, the location of the maximum mass systematically shifts towards higher central densities, underscoring the enhanced stiffness contributed by higher-curvature terms. This response accords with earlier detailed investigations of neutron stars employing realistic equations of state within the regularized four-dimensional Einstein-Gauss-Bonnet framework \cite{2025PhRvD.111f4071S,Gammon:2023uss}, and is in agreement with the established turning-point stability criteria for relativistic stars \cite{1965gtgc.book.....H,1971reas.book.....Z}.

\section{Discussion and final remarks}
\label{sec:disc}

\smallskip\noindent

Our investigation of compact star configurations within regularized four-dimensional Einstein-Gauss-Bonnet gravity reveals systematic deviations from GR that remain consistent with multi-messenger observations. The analysis employed a color-flavor locked quark matter equation of state, enabling semi-analytical control whilst capturing essential physics of deconfined matter at extreme densities.

\smallskip\noindent

The main results indicate that Gauss-Bonnet corrections with coupling values in the range $\alpha=(5-10)~(km)^2$ lead to stellar sequences with three distinct features. First, the maximum stellar masses increase by about $(9.7-19.9) \%$ compared to the predictions of GR, thereby allowing for the existence of heavier pulsars within observational limits. Second, the radii expand systematically by $(1.3-2.4) \%$, resulting in somewhat larger compactness factors (see Table 1). Third, the resulting mass-radius curves remain fully compatible with the latest multimessenger constraints, including pulsar timing, NICER radius measurements, and gravitational-wave observations \cite{Miller:2021qha,Riley:2021pdl,LIGOScientific:2020zkf,2021JCAP...05..024D,2025PhRvD.111f4071S}.

\smallskip\noindent

These deviations originate from an effective stiffening of the equation of state, produced by the higher-order curvature contributions in the modified field equations. In particular, the factor $\left(1+2\alpha(1-\mathrm{e}^{-2\Psi})/r^{2}\right)$ in equations \eqref{FEq1} and \eqref{FEq2} reduces the effective gravitational coupling in the high-curvature regime near the stellar core. This reduction naturally generates less compact stellar models while preserving stability against radial oscillations, a result confirmed by the persistence of the turning-point behavior in the $M(\rho_c)$ relation \cite{1965gtgc.book.....H,1971reas.book.....Z}.While Reference 
\cite{2025PhRvD.111f4071S}	 shows that the fundamental Harrison-Zel'dovich stability criterion is preserved (with transitions still occurring at maximum mass points), the Gauss-Bonnet corrections do modify the stability landscape by shifting where these transitions occur and, in some cases, allowing configurations to re-approach stability near black hole limits.

\smallskip\noindent

The CFL EoS serves as a practical tool for probing modifications to gravity in compact stars \cite{Panotopoulos:2019wsy,Lopes:2019psm}. Its explicit dependence on the color superconducting gap $\Delta$ and the strange quark mass $m_s$, as presented in equation~\eqref{alfa}, enables a quantitative connection between microscopic physics and astrophysical observables~\cite{lugones2002,2017PhRvC..95b5808F}. Astrophysical investigations, especially in the $4$DEGB framework, demonstrate that models with $\Delta \simeq 100\,\mathrm{MeV}$ and bag constant $B^{1/4} \simeq 150\,\mathrm{MeV}$ are able to produce compact star sequences with gravitational masses exceeding $2\,M_\odot$, while their radii remain compatible with the most recent NICER measurements~\cite{2024ApJ...971L..19R,2024ApJ...974..295D,2021ApJ...918L..28M,2021ApJ...918L..27R}. Notably, recent studies have placed 95\% upper limits on the CFL pairing gap that are consistent with the values employed in this work~\cite{2024PhRvL.132w2701K}.

\smallskip\noindent

Assessment relative to complementary literature confirms the robustness of these findings. Analyses incorporating hadronic equations of state consistently report a 5--15 per cent increase in maximum mass for similar choices of the Gauss-Bonnet coupling~\cite{2021JCAP...05..024D,2025PhRvD.111f4071S,2024arXiv241203348R}. Work on electrically charged quark stars in the $4$DEGB context verifies that similar trends are observed, underlining the relative independence from the equation of state and providing additional latitude for massive objects~\cite{2022EPJC...82..180P}. Studies of rotating compact stars~\cite{2025PhRvD.111f4071S,2011PhRvD..84j4035P} further show that these qualitative modifications persist even when rotational corrections are introduced. Altogether, such convergence across independent theoretical approaches and parameter regimes, supported by up-to-date multimessenger constraints, corroborates the conclusion that higher-curvature gravity terms yield a well-motivated extension to general relativity in the astrophysical regime.

\smallskip\noindent

Beyond the results presented here, several open directions remain to be explored. A natural extension involves detailed calculations of tidal deformabilities in order to directly confront gravitational-wave measurements from binary neutron star mergers~\cite{2018PhRvL.121i1102D,2017PhRvL.119p1101A}. Radial oscillation spectra provide an additional diagnostic tool, yielding asteroseismic constraints through the quasi-periodic oscillations observed in magnetar flares~\cite{2020PhRvD.101f3025S}. Furthermore, the implementation of realistic hybrid equations of state including phase transitions would allow one to assess whether Gauss--Bonnet corrections modify the stability of twin-star branches and influence the properties of high-mass compact configurations~\cite{2024arXiv241203348R,2019PhRvL.122f1102B}.

\smallskip\noindent

The astrophysical consequences of these findings extend well beyond the properties of isolated compact stars. A confirmed systematic reduction in stellar compactness would have measurable effects on kilonova light curves, influence the pathways and yields of heavy-element production during r-process nucleosynthesis, and modify the gravitational-wave signatures produced by binary inspiral events \cite{2018ApJ...852L..29R,2017PhRvL.119p1101A}. Furthermore, a moderate increase in the maximum allowable mass could permit the existence of more massive or exotic compact objects, all while preserving consistency with empirical bounds from nuclear physics at lower densities \cite{2021JCAP...05..024D,2025PhRvD.111f4071S}.

\smallskip\noindent

In summary, regularized four-dimensional Einstein-Gauss-Bonnet gravity provides a consistent and testable framework for describing strong-field stellar structure. The theory predicts measurable deviations from general relativity that remain compatible with present multimessenger constraints. As observational precision improves, particularly in the domains of gravitational-wave astronomy and X-ray timing, these predictions will be increasingly tested, offering the potential to uncover signatures of modified gravity in the most extreme astrophysical environments.

\section*{Acknowledgments}
\smallskip\noindent
I.~L. thanks the Fundação para a Ciência e Tecnologia (FCT), Portugal,  for the financial support to the Center for Astrophysics and Gravitation (CENTRA/IST/ULisboa) through grant No. UID/PRR/00099/2025 (https://doi.org/10.54499/UID/PRR/00099/2025) and grant No. UID/00099/2025 (https://doi.org/10.54499/UID/00099/2025).
A.~R. would like to express his gratitude to Silesian University in Opava, Czech Republic, for their financial support.
The creation of this article was supported by the grant program Vouchers for Universities in the Moravian-Silesian Region (registration number $CZ.10.03.01/00/23\_042/0000390$).

\bibliographystyle{utphys}
\bibliography{Library2}

\end{document}